\documentclass[manuscript]{aastex}

\newcommand{\solar}{\ifmmode_{\mathord\odot}\else$_{\mathord\odot}$\fi}
\newcommand{\gray}{$\gamma$-ray\ } \newcommand{\grays}{$\gamma$-rays\ }
\newcommand{\mic}{$\mu$m\ } 
\newcommand{\etal}{et al.\ }

\begin{document}

\title{An Empirical Determination of the Intergalactic Background Light from UV to FIR Wavelengths Using FIR Deep Galaxy Surveys and the Gamma-ray Opacity of the Universe}

\author{Floyd W. Stecker}  
\affil{Astrophysics Science Division, NASA/Goddard  Space   Flight  Center}
\authoraddr{Greenbelt, MD 20771; Floyd.W.Stecker@nasa.gov}
\affil{Department of Physics  and Astronomy, University of California,
Los  Angeles}
\author{Sean T. Scully}  
\affil{Department  of Physics,  James  Madison
University} 
\authoraddr{Harrisonburg, VA 22807; scullyst@jmu.edu}  
\author{Matthew A. Malkan}  
\affil{Department  of Physics  and  Astronomy, University  of
California,   Los  Angeles}  
\authoraddr{Los   Angeles,  CA90095-1547;
malkan@astro.ucla.edu}

\begin{abstract}
We have previously calculated the intergalactic background light (IBL) as a function of redshift from the Lyman limit in the far ultraviolet to a wavelength of 5 $\mu$m near infrared range, based purely on data from deep galaxy surveys. Here we utilize similar methods to determine the mid- and far-infrared IBL from 5 $\mu$m to 850 $\mu$m. Our approach enables us to constrain the range of photon densities, by determining the uncertainties in observationally determined luminosity densities and spectral gradients. By also including the effect of the 2.7 K cosmic background photons, we determine upper and lower limits on the opacity of the universe to \grays up to PeV energies within a 68\% confidence band.

Our direct results on the IBL are consistent with those from complimentary \gray analyses using observations from the {\it Fermi} \gray space telescope and the H.E.S.S. air \v{C}erenkov telescope.
Thus, we find no evidence of previously suggested processes for the modification of \gray spectra other than that of absorption by pair production alone.

\end{abstract}

\subjectheadings{diffuse radiation -- galaxies: observations -- gamma-rays: theory} 

\section{Introduction}

We have previously employed a {\it fully empirical} approach to calculating the intergalactic background light (IBL) to wavelengths up to 5 $\mu$m by using observational data from deep galaxy surveys (Stecker, Malkan \& Scully 2012 (hereafter SMS12); and Scully, Malkan \& Stecker 2014 (hereafter SMS14)); for a similar approach, see also Helgason \& Kashlinsky (2012) and Khaire \& Srianand (2015). Here we extend our previous results from SMS12 and SMS14, determining both the intergalactic background light (IBL) at longer wavelengths out to 850 $\mu$m  and the subsequent $\gamma$-ray opacity of the Universe out to multi-TeV energies. The spectra of astrophysical sources of such high energy \grays are being studied by ground-based air \v{C}erenkov telescopes and linked air \v{C}erenkov telescope arrays. 

We accomplish the goals of this paper by using very recent deep galaxy survey data at far infrared wavelengths where galaxy emission is produced by dust re-radiation rather than starlight. We also include the \gray opacity from photons of the 2.7 K cosmic background radiation (CBR). This enables us to extend our calculations of the \gray opacity of the Universe to energies much greater than a TeV. 

Observations at wavelengths greater than 24 $\mu$m, have been covered by the Multiband Imaging Photometer on the {\it Spitzer} space telescope (MIPS), now being dramatically advanced by the availability of data from the Photoconductor Array Camera and Spectrometer (PACS) and Spectral Photometric Imaging Receiver (SPIRE) instruments on the {\it Herschel} space telescope, the {\it Planck} space telescope, and by ground-based observations from the Atacama Large Millimeter Array (ALMA) and observations from the Balloon-borne Large Aperture Submillimeter Telescope (BLAST).  By using only empirical data rather that modeling, our approach is, by definition, {\it model independent}. We use published luminosity functions and interpolations of luminosity densities between observed wavebands, observations from high-redshift galaxy surveys being now sufficient for this purpose. 
 
Most past approaches to  determining the IBL and its present-day spectral energy distribution, referred to as the "extralgalactic background light (EBL)", require assumptions about how the galaxy luminosity functions evolve ({\it e.g.} Malkan \& Stecker 1998, 2001; Kneiske, Mannheim \& Hartmann, 2002: Stecker, Malkan \& Scully 2006 (SMS06); Franceschini et al. 2008; Finke, Razzaque, \& Dermer, 2010; Kneiske \& Dole 2010). Other approaches make use of semi-analytic models that require various assumptions regarding galaxy evolution, stellar population synthesis modeling, or star formation rates the properties of and dust attenuation, particularly for redshifts greater than 1 ({\it e.g.}, Salamon \& Stecker 1998; Gilmore et al. 2009; Somerville et al. 2012; Dominguez et al. 2011; Inoue et al. 2013; Driver et al. 2016). In contrast, our observationally-based approach is superior to model-based methods, since it enables a determination of both the IBL {\it and its observational uncertainties}. This is because we use observationally determined errors. Approaches that rely on modeling cannot determine such uncertainties. 

Observational studies of blazar $\gamma$-ray spectra have also been used to probe the IBL (Ackermann et al. 2012; Abramowski et al. 2013; Biteau \& Williams 2015) through its opacity effect caused by electron-positron pair production. This approach was originally suggested by Stecker, De Jager \& Salamon (1992) when the infrared EBL was unknown.  Our present method of determining the expected \gray opacity of the Universe by first using observational data to determine the IBL and then calculating its effect on \gray spectra of extragalactic sources complements the technique of using direct \gray observations. This is because the later approach requires a thorough knowledge of the intrinsic (unabsorbed) emission spectra of the \gray sources. This requirement introduces unavoidable uncertainties.

Our final results give the \gray opacity as a function of energy and redshift to within a 68\% confidence band that is based on observational data. This opacity is solely caused by pair-production interactions of IBL photons with extragalactic \grays. Thus, a direct comparison of this effect with \gray spectra of extragalactic sources enables an assessment of possible additional spectral modifications. One such possible spectral modification has been suggested to be caused by secondary \gray production from cosmic-ray interactions along the line of sight to the source (Essey \& Kusenko 2014). Another such modification might be caused by photon-axion oscillations during propagation from the source to the Earth ({\it e.g.}, De Angelis et al. 2007; Mayer \& Horns 2013). 

In Section 2 we give the details of our calculations of upper and lower limits on IR galaxy luminosity densities for wavelengths between 5 \mic and 850 \mic as a function of redshift, paying particular attention to the effects of the emission spectra of polycyclic aromatic hydrocarbons (PAHs). In Section 3 we calculate the $z = 0$ EBL based on the results from Section 2 and compare it with both other work and observational limits.  In Section 4 we compute the resulting opacity of the universe to \grays out to a redshift of z = 5, including the total opacity from $\gamma + \gamma \rightarrow e^+ + e^-$ interactions with both IBL and thermal cosmic background radiation (CBR) photons. In our conclusion section, Section 5, we compare our results with those obtained by other methods in some of the papers mentioned above. We also discuss the implications of our results.

\section{Calculating the Infrared IBL}

We have previously calculated the intergalactic background light (IBL) as a function of redshift in the far ultraviolet to near infrared range, based purely on data from deep galaxy surveys (SMS12; SMS14). Here we utilize similar methods to extend the calculation into the mid- and far infrared IBL out to a wavelength of 850 $\mu$m.

\subsection{Determining the Luminosity Densities from Empirical Luminosity Functions}

In our previous work the observationally tolerated ranges of photon densities were determined from the luminosity densities (LDs), $\rho_{L_{\nu}}$, themselves, with errors provided by the various authors at wavelengths ranging from the far UV to the 5 \mic wavelength in the near IR. The LDs are computed by integrating fits to the observationally determined luminosity functions:
\begin{equation}
 \rho_{L_{\nu}} = \int_{L_{min}}^{L_{max}} dL_{\nu} \, L_{\nu} \Phi(L_{\nu};z)
\label{lumdens}
\end{equation}
Cases where authors did not directly compute the LDs were excluded because properly estimating their error requires knowledge of the covariance of the errors in the fit parameters of the luminosity functions (LFs) and also knowledge of any observational biases. To provide comprehensive redshift coverage of the LDs, SMS12 and SMS14 made use of continuum colors between the wavelength bands to fill in any gaps. 
 
At wavelengths greater than 5 \mic very few studies provide determinations of luminosity densities (LDs) directly, as most authors are more concerned with calculating the {\it total} IR density, integrated over all wavelengths. This is because the total IR density is an observable that is correlated with the star formation rate.  Therefore, in this paper we used observer-given analytic fits to the LFs at various wavelengths. When those fits were not provided, we ourselves fit the observed infrared galaxy LFs at various wavelengths and redshifts and then use equation (\ref{lumdens}) to obtain the LDs. 

Galaxy LFs typically have characteristic shapes that are flatter at lower luminosities but fall off more steeply at the highest luminosities. However, at wavelengths greater than 5 \mic the Schechter function, commonly used at optical wavelengths, does not provide a good fit to the observed LFs, as its exponential decrease falls off too quickly at the bright end compared with the measured LFs. Most observers instead fit their LFs to a broken power-law function that describes the data better (e.g., Marchetti et al. 2016). The transition region between the power law that holds at low luminosities and that holding at high luminosities is usually referred to as the {\it knee} of the LF. The luminosity at the knee is usually designated as $L_*$. Most of the luminosity density from an LF is produced by galaxies with luminosities in the vicinity of the knee; much fainter galaxies do not contribute much to the LD and the much brighter ones are quite rare. Thus, it is critical to determine the location of the knee in order to estimate the LDs with any accuracy. 

For the cases where we were required to determine our own fits, we chose a double power-law fitting function of the form:
\begin{equation}
\Phi (L) = \frac{c}{{L_*\left( {{{\left( {\frac{L}{L_*}} \right)}^{a}} + {{\left( {\frac{L}{L_*}} \right)}^{b}}} \right)}}
\label{lumfunc}
\end{equation}
Here, the parameters $a$ and $b$ are the indices of the power-law fits in the low luminosity and high luminosity ranges respectively. The overall normalization of the LF is given by the parameter $c$, which has the dimension of luminosity per Mpc$^3$dex. To compute the LDs, we integrate equation (\ref{lumdens}) over the galaxy luminosities between 4 $\times$ 10$^7$ L$_{\solar}$ and 10$^{14}$ L$_{\solar}$. 

In order to accurately compute the errors in the LDs as derived from the fit parameters, which is essential for our calculation, we must do so in a way that includes terms involving the off-diagonal elements of the error matrix of the fits so as to account for covariance. In cases where the authors provided there own fits, we have therefore re-derived the fits of the LFs to generate the error matrix, retaining the same choice of fitting function and fixed parameters, if given. 

Our goal, as in SMS12 and SSM14, was to compute the observationally tolerated ranges of luminosity densities. This requires that we represent the error on these quantities as best we can. The statistical error is determined by properly propagating the fit parameter errors accounting for covariance, lest we overestimate the error. To compute the total error on each LD, we further added in quadrature an additional systematic error that accounts for cosmic variance. We compute cosmic variance based on the field sizes of the individual studies. For the {\it {\it AKARI} Wide Field IR Survey Explorer (WISE)} and {\it GOODS} fields this value is typically of order of 10\%. 
In determining the LDs, for consistency, all observational LFs are scaled to a Hubble parameter of $h = 0.7$

Our new additions to our observationally determined LDs extend our coverage of rest frame galaxy photon production from the near IR to 850 \mic in the far IR, with enough determinations at each wavelength band to span the redshift range $0 \le z \le 2 - 3$.  Using published results derived from observations by the {\it Spitzer} and {\it Herschel} space telescopes, sufficient redshift coverage was found for wavelength bands of $8, 12, 15, 24, 35, 60, 90,$ and $250$ \mic.  

There were two cases where we were required to combine LFs from different observational studies in order to provide enough coverage to discriminate the location of $L_*$ particularly for redshifts greater than $1$. At 12 $\mu$m we combined the results of Perez-Gonzalez \etal (2005) and Rodighiero \etal (2010) to compute LDs in redshift bins centered on redshifts of $1.2, 1.6,$ and $2.0$. In order to achieve sufficient redshift coverage at $90$ $\mu$m we combined some of the higher redshift $100$ \mic data from Lapi \etal (2011) with that of the $90$ \mic data from Gruppioni et al. (2013). This gains us additional coverage at redshifts of $1.4, 2.2,$ and $3.0$.

\begin{figure}
\includegraphics[width = 7.0in]{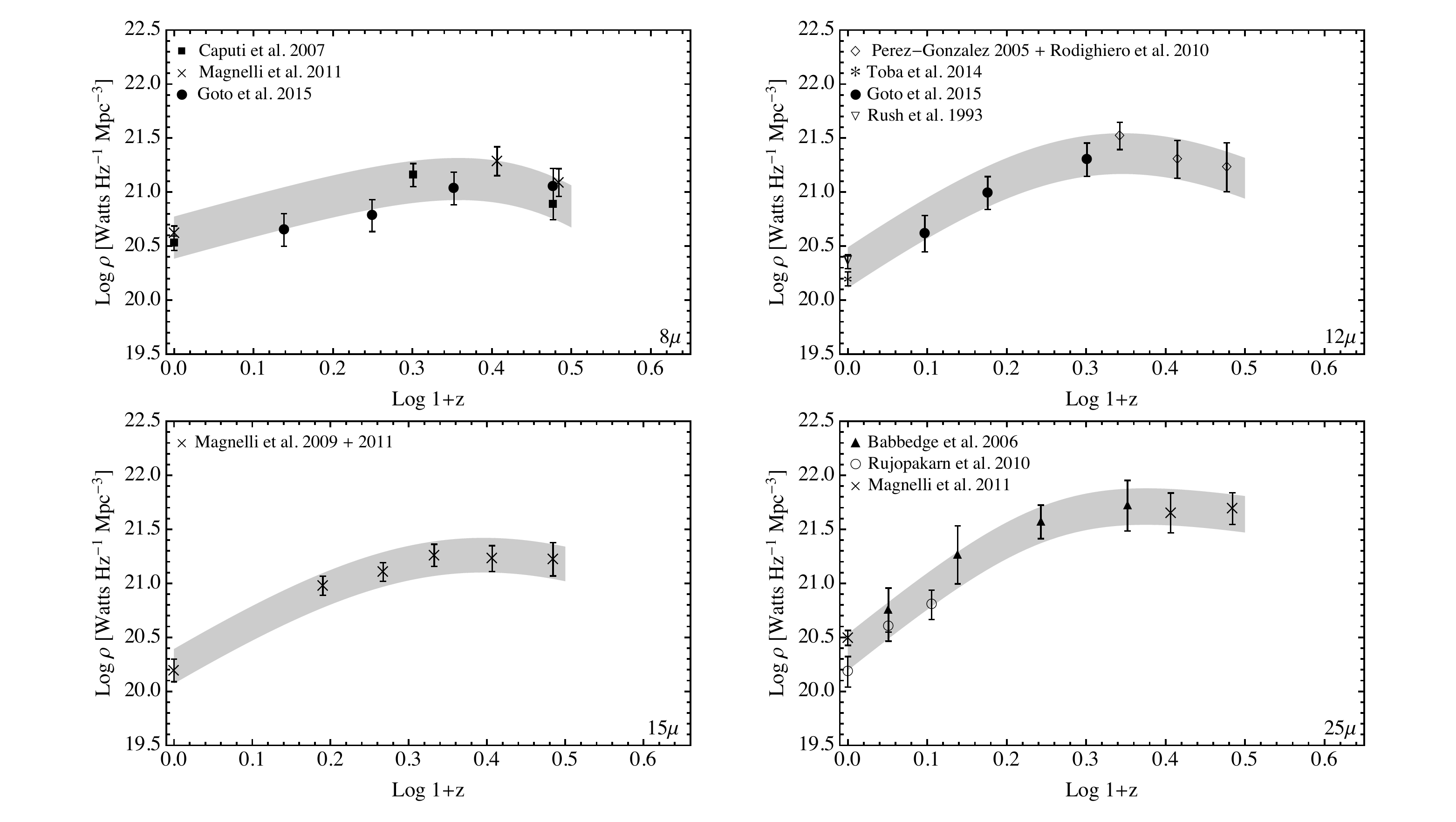}
\caption{The luminosity densities for 8, 12, 15, and 24 $\mu$m wavebands. Data are from several sources. Some {\it Spitzer} data from all 4 wavebands are from Rodighiero et al. (2010). {\it AKARI} 8 and 12 \mic data are from Goto et al. (2015), {\it Spitzer} data at 8 \mic are from Huang et al. (2007) and Caputi et al. (2007). {\it Spitzer} data at 8 and 24 \mic are from Babbedge et al. (2006). {\it Spitzer} data at 15 \mic are from Le Floc'h et al. (2005). {\it Spitzer} data at 24 \mic are from P\'{e}rez-Gonz\'{a}lez et al. (2005). 
The grey shading represents the 68\% confidence bands (see text).}
\label{lumdens1}
\end{figure}

\begin{figure}
\includegraphics[width = 7.0in]{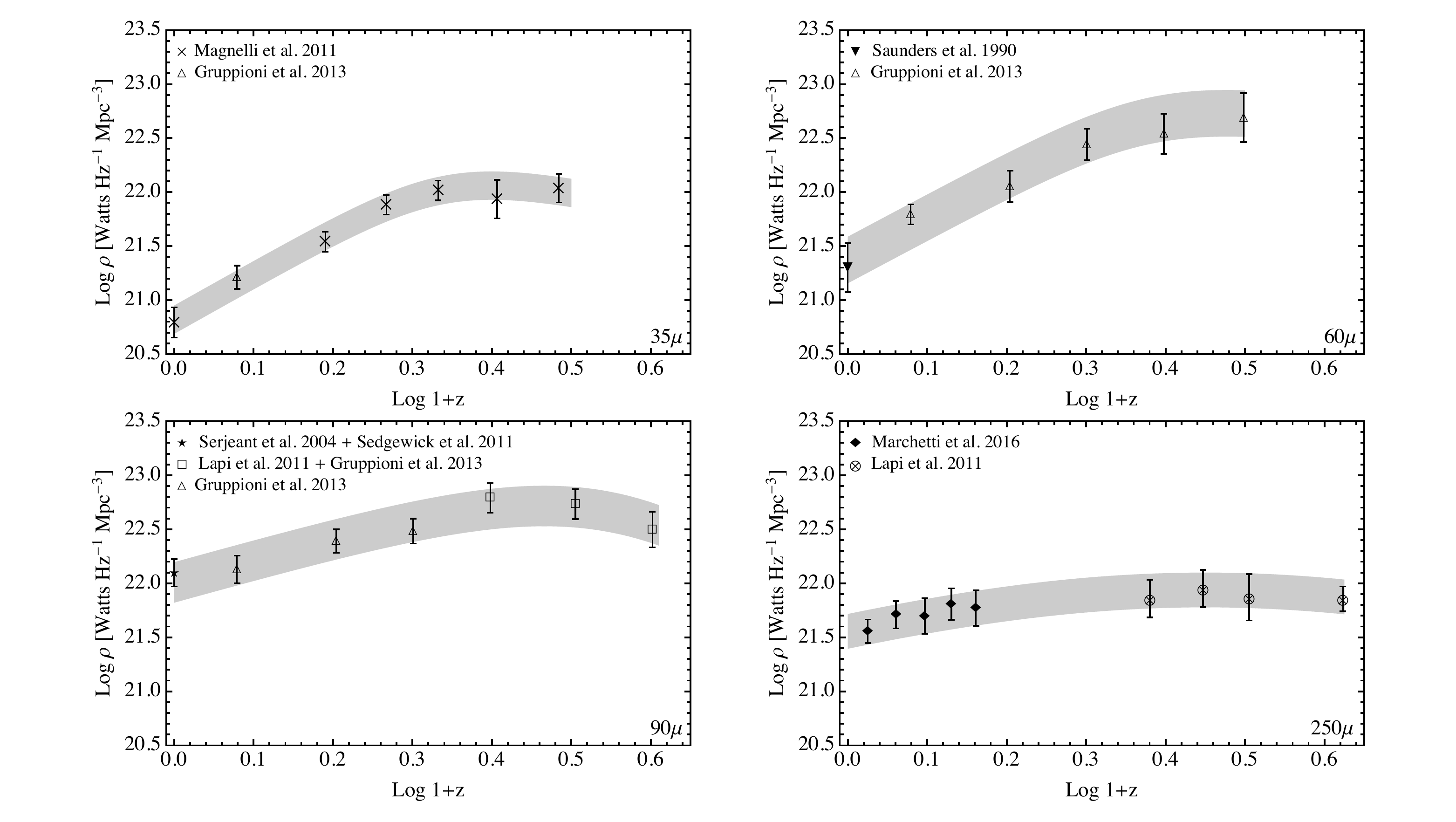}\caption{The luminosity densities for the 35, 60, 90, and 250 $\mu$m wavebands.The grey shading represents the 68\% confidence bands. {\it Herschel} data at 35 60, and 90 \mic are from Gruppioni et al. (2013). Also 100 \mic data from Lapi et al (2011) are plotted on the 90 \mic graph. {\it Herschel} data at 250 \mic are from Eales et al. (2010), Dye et al. (2010) and Smith et al. (2012) (see text).}
\label{lumdens2}
\end{figure}

At 160, 350, 500, and 850 \mic, LF data only exists for the very nearby redshifts.  We therefore assumed that their redshift evolution closely follows that of the 250 \mic band (Lapi et al. 2011; Marchetti et al. 2016). We use local LDs calculated in those bands as a normalization to this evolution. This assumption is justified because the emission in this wavelength region is dominated by warm dust. At 160 \mic we used the local LF given be Patel et al. (2013). At 350 and 500 \mic, we used the combined local LF data of Marchetti \etal (2016) from {\it Herschel}/SPIRE, the {\it Herschel}/SPIRE estimate of Vaccari \etal (2010), and the {\it Planck} satellite data from Negrello \etal (2013). At 850 $\mu$m we computed the local LD from the LF provided by Negrello \etal (2013). Figures \ref{lumdens1} and \ref{lumdens2} show the resulting derived values for $\rho_{L_{\lambda}}(z)$ and their errors together with the observationally determined $\pm$ 1 $\sigma$ confidence bands for the 8, 12, 15, 24 \mic bands and those of 35, 60, 90, and 250 \mic respectively.
 
Our calculations of LDs at mid-IR to far-IR wavelengths presented here is an extension of the work done in SMS12 and SSM14.  Thus, the results given in those papers for wavelengths less than 5 \mic is almost unchanged from the results presented here.  However, we have updated the far UV calculations to include the more recent work from Bouwens \etal (2015) for LDs in the redshift range of $4$ to $7$.  Even though the shape of the far UV band can affect other bands when filling in for redshift gaps using colors, the overall calculation yields results that are qualitatively the same as those presented in SMS12 and SSM14, as the newer data do not significantly change the general trend.

In order to place 68\% upper and lower limits by using observational data on $\rho_{L_{\nu}}$ we make as few assumptions about the luminosity density evolution as possible. In SMS12 and SMS14 we utilized a robust rational fitting function in the form of a broken power-law dependent on $(1 + z)$ in order to generate the confidence bands. We take the 68\% confidence ranges of these fits in each waveband as the $\pm$1 $\sigma$ confidence bands for the LDs. At redshifts beyond the redshift at the peak of star formation, the LDs decline with redshift. In accord with recent studies of the evolution of rest frame LFs in the UV that trace the star formation rate (Finkelstein et al. 2015), we conservatively assume upper and lower limit power-law functions in redshift to represent the rate of this decline as the highest redshifts. In the upper limit case, we assume a decline proportional to $(1+z)^{-2}$. In the lower limit case, we adopt a steeper decline proportional to $(1+z)^{-4}$.  These assumptions have almost no impact on the derivation of the opacity confidence bands that we determined. Figures \ref{lumdens1} and \ref{lumdens2} show our results for LDs as
a function of redshift at various wavelengths.

\subsection{Taking account of PAH emission}

At wavelengths greater than $\sim$20 \mic the LDs between the bands can be determined by smoothly interpolating between our observationally based LDs at specific wavelengths, since the spectral energy distributions (SEDs) from galaxies in this wavelength range are smooth modified blackbody spectra produced by dust re-radiation. However, in the 5 -- 20 \mic range the situation is more complex. 

In star-forming galaxies, the spectra between 7 and 13 $\mu$m are dominated primarily by PAH emission. These PAH molecules are found in very small dust grains in intergalactic media. They absorb the UV photons emitted by hot young $O$ and $B$ stars and reemit them in molecular emission bands in the mid-IR. They are thus a strong signature of active star formation in galaxies (Peeters, Spoon \& Tielens 2004). In this regard, we note that luminous star forming galaxies at higher redshift have more prominent PAH emission features. The importance of PAH features in the mid-IR at redshifts $0.5 \le z \le 2.5$ has been shown by Lagache et al. (2005).

\begin{figure}
\includegraphics[width = 6.5in]{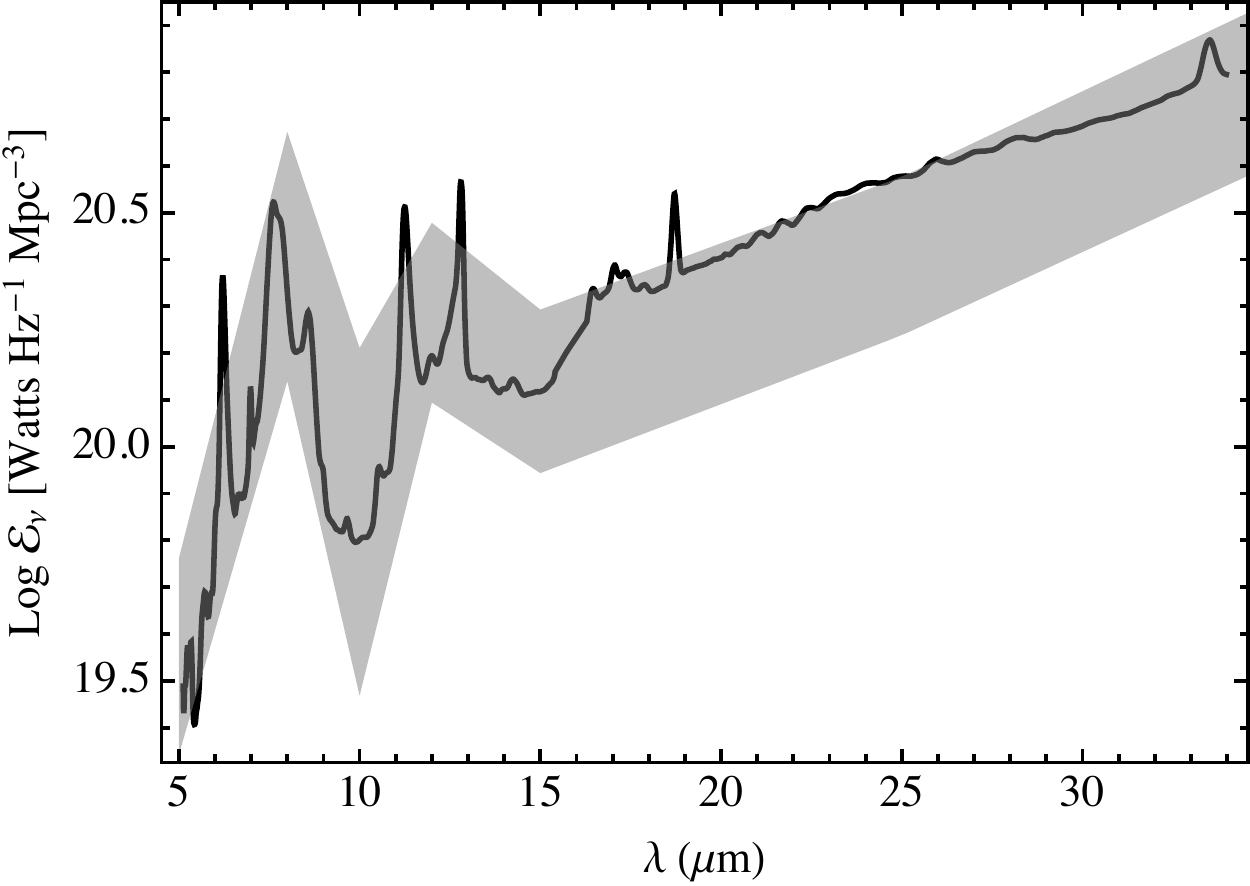}
\caption{Average low redshift galaxy SED at MIR wavelengths based on the Class 1C SED of Spoon et al. (2007) normalized to our low redshift LDs from Figures \ref{lumdens1} and \ref{lumdens2}, as indicated by the shaded region.} 
\label{PAH}
\end{figure}

The average SED of nearby star-forming galaxies (from Spoon et al. 2007, see also Smith et al. 2007) is shown in Figure \ref{PAH} normalized to our best fit low redshift LD confidence band. One can see that there is a relative "valley" between 9 and 11 $\mu$m.  A simple direct interpolation between our 8 and 12 $\mu$m bands would therefore obtain an incorrectly high value for the LD in this wavelength range.  Since we do not have wavelength coverage in this regime, we take this feature into account by lowering our interpolated LDs by a factor of 3 for the upper limit and a factor of 5 for the lower limit at 10 \mic. The factor of 5 is chosen as a lower limit based on the difference between the peak at 8 \mic and the depth of the valley near 10 \mic from the SED.  For the upper limit, we relax the depth of this feature because, while star forming galaxies make up the bulk of the IBL, contributions from other galaxy types have a less pronounced PAH feature. Since, at high redshifts, PAH emission correlates with the star formation rate (Shipley et al. 2016), we assume that our relative PAH shape factors of 3 and 5 coevolve with redshift.

\section{The Extragalactic Background Light}

At wavelengths above $\sim$ 100 \mic the FIRAS and DIRBE instruments aboard the {\it Cosmic Background Explorer} (COBE) have measured the EBL (Fixsen et al. 1998; Lagache et al. 1999).
This diffuse background has now been in large part resolved by recent galaxy count studies using the {\it Herschel} telescope (Berta et al. 2010; B\'{e}thermin et al. 2012; Viero et al. 2015)
along with ground based studies using {\it ALMA} (Fujimoto et al. 2016) and {\it BLAST} (Marsden et al. 2011). These more recent studies strongly support the COBE results.

However, there are no direct measurements of the EBL in the infrared range below $\sim$ 100 \mic owing to the predominance of the foreground radiation of Zodiacal light from interplanetary dust re-radiation. The flux of Zodiacal light is approximately two orders-of-magnitude larger than the EBL flux (Spiesman et al. 1995). In this region only lower limits obtained from galaxy counts exist.

\begin{figure}
\includegraphics[width = 6.5in]{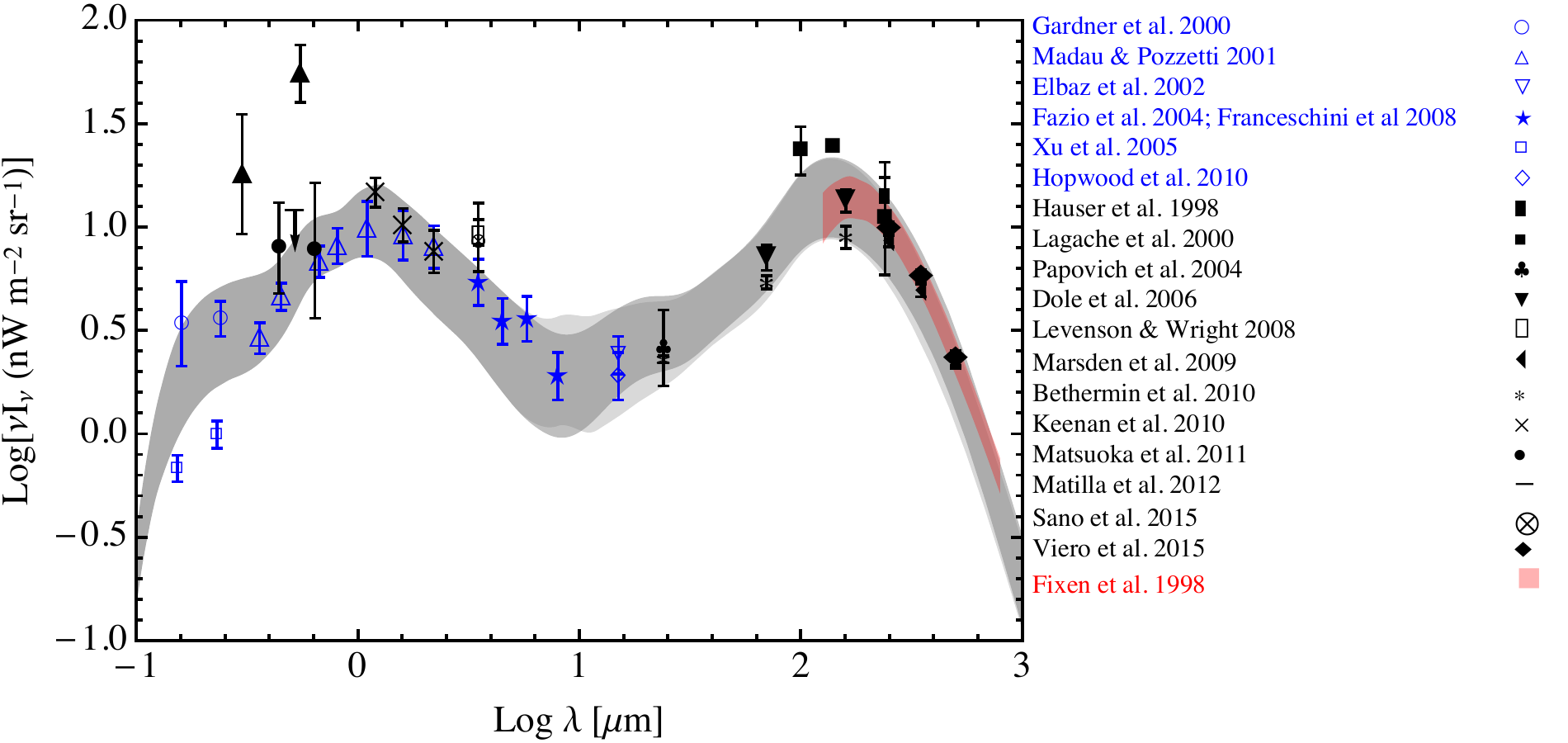}
\caption{Our spectral energy distribution of the EBL together with empirical data based on our mid-IR LDs and far-IR LDs and the results of SMS12 and SMS14. The light shaded area between $\sim$10 \mic and $\sim$30 \mic indicates the maximum effect of the PAH bands (see Sect. 2.2). The lower limits from galaxy counts are shown in blue; direct measurements and extrapolations from galaxy counts in the mid-IR are shown in black. References for the empirical data before 2012 are given by Lagache et al. (2005) and Dwek \& Krennrich (2013). A 3.5 \mic point is from Sano et al. (2016). The red shaded area is based on the {\it COBE-FIRAS} results of Fixsen et al. (1998) with limits described by modified black body spectra.} 
\label{ebl-data}
\end{figure}

Using our calculations of the LDs in the mid-IR and far-IR as a function of wavelength, we have constructed a 68\% confidence band for the spectral energy distribution of the cosmic diffuse infrared background light (the IBL at $z = 0$). Figure \ref{ebl-data} shows our new results, combined with our previous results at shorter wavelengths, taken from SMS12 and SMS14. The light shaded band shows the maximum effect of PAH emission. It can be seen that taking account of the details of the PAH spectrum does not significantly affect our EBL results. Figure \ref{ebl-data} also shows the observational lower limits on the EBL obtained from galaxy counts (in blue), extrapolations of mid-IR galaxy counts from {\it Spitzer}, and direct measurements (in black).

\section{The Optical Depth from $\gamma + \gamma \rightarrow e^+ + e^-$ Interactions with IBL  and 2.7 K CBR Photons}

The co-moving radiation energy density for wavelength $\lambda$ at redshift $z$, $u_{\nu}(z)$, where $\nu = c/\lambda$, is the time integral of the co-moving luminosity density ${\rho}_{\nu}(z)$,
\begin{equation} 
\label{u1}
u_{\nu}(z)=
\int_{z}^{z_{\rm max}}dz^{\prime}\,{\rho}_{\nu^{\prime}}(z^{\prime})
\frac{dt}{dz}{(z^{\prime})},
\end{equation}

\noindent where $\nu^{\prime}=\nu(1+z^{\prime})/(1+z)$ and $z_{\rm max}$ is the
redshift corresponding to initial galaxy formation 
(Salamon \& Stecker 1998), and
\begin{equation}
\frac{dt}{dz}{(z)} = {[H_{0}(1+z)\sqrt{\Omega_{\Lambda} + \Omega_{m}(1+z)^3}}]^{-1},
\label{cosmology}
\end{equation}

\noindent with $\Omega_{\Lambda} = 0.72$ and $\Omega_{m} = 0.28$.

\begin{figure}
\includegraphics[width = 6.5in]{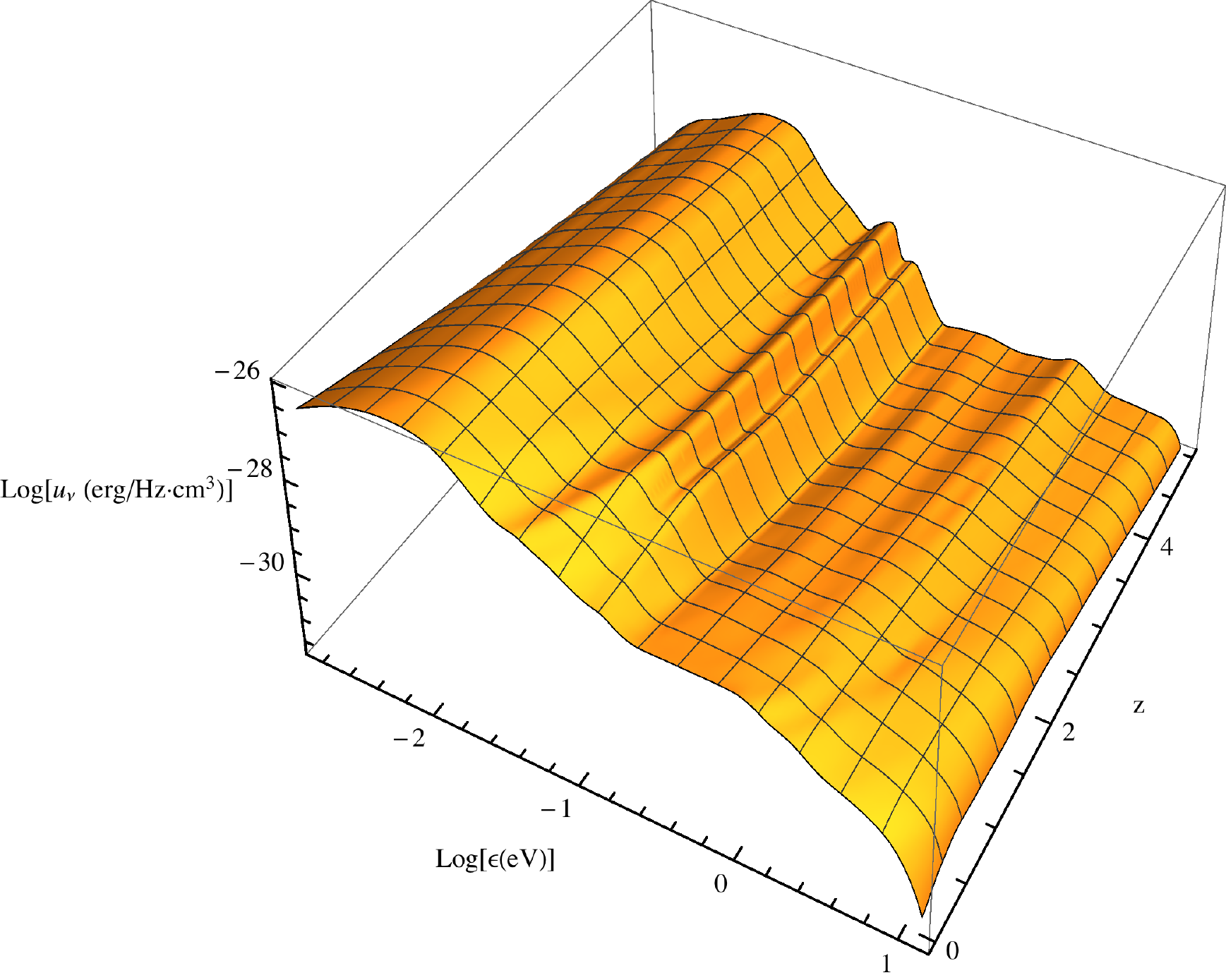}
\caption{Upper limit envelope on the co-moving energy density as a function of energy and redshift.}
\label{up}
\end{figure}

\begin{figure}
\includegraphics[width = 6.5in]{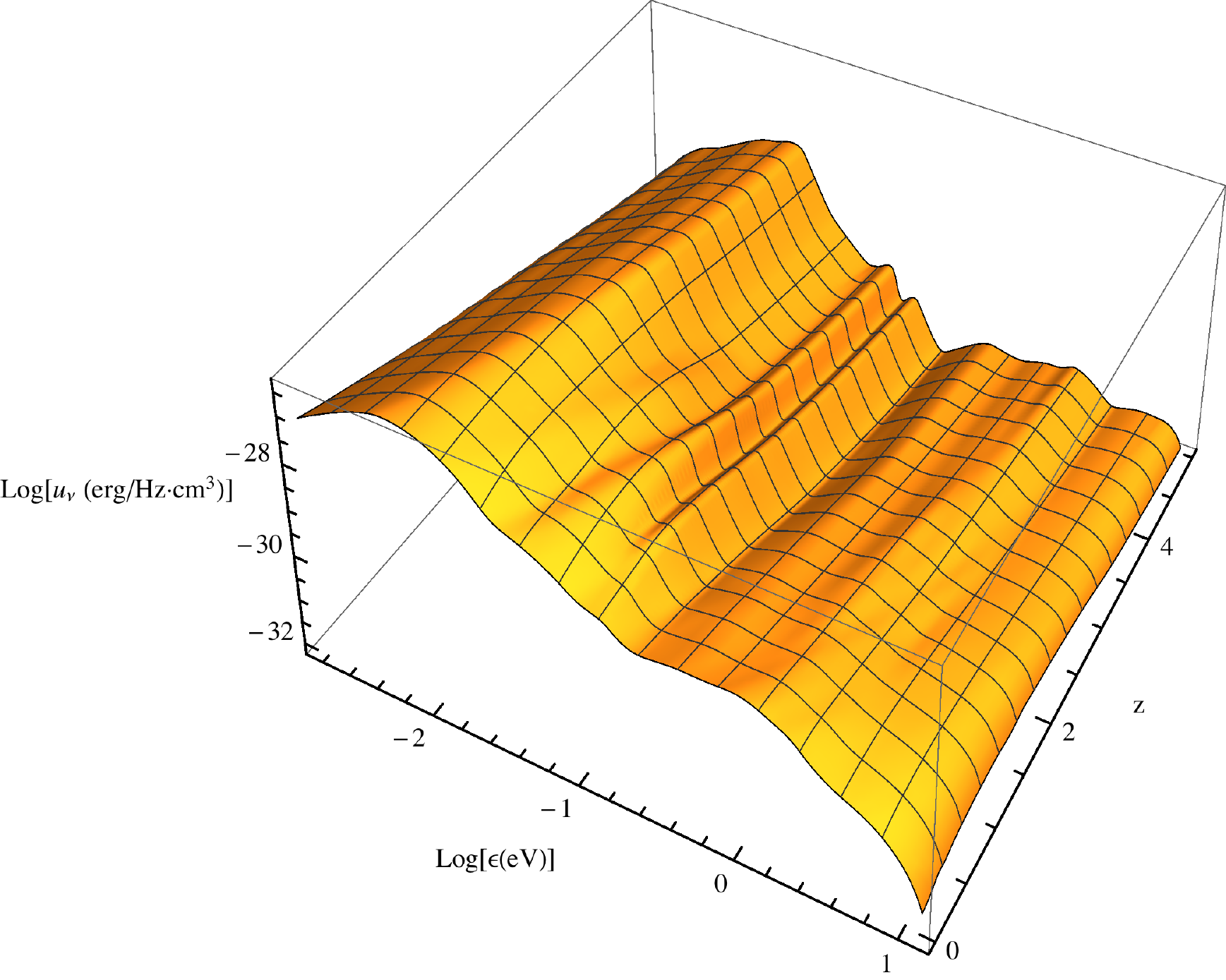}
\caption{Lower limit envelope on the co-moving energy density as a function of energy and redshift.}
\label{lo}
\end{figure}

The upper and lower limits on our co-moving energy densities, derived using equation (\ref{u1}),
are shown in Figures \ref{up} and \ref{lo}.

In calculating the \gray opacities we use the relations for the photon energy, $\epsilon_\nu = h\nu$, and the photon density, $n_\nu = \rho_\nu /\epsilon$.

The cross section for photon-photon annihilation to electron-positron pairs was first calculated by Breit \& Wheeler (1934) as a solid result of quantum electrodynamics.  
The threshold for this interaction follows from the Lorentz invariance of the square of the four-momentum vector. This reduces to the square of the threshold energy in the c.m.s., $s$, that  is necessary to produce twice the electron rest mass:  
\begin{equation}
s = 2\epsilon E_{\gamma} (1-\cos\theta) = 4m_{e}^2
\label{s}
\end{equation}

The Lorentz invariant quantity given by $s$ has been determined to hold to better than one part in $10^{15}$ (Stecker \& Glashow 2001;
Jacobson, Liberati, Mattingly \& Stecker 2004).

The optical
depth for \grays caused by electron-positron pair production 
interactions with photons of the stellar radiation
background is given by 

\begin{equation} 
\label{tau}
\tau(E_{0},z_{e})=c\int_{0}^{z_{e}}dz\,\frac{dt}{dz}\int_{0}^{2}
dx\,\frac{x}{2}\int_{0}^{\infty}d\nu\,(1+z)^{3}\left[\frac{u_{\nu}(z)}
{h\nu}\right]\sigma_{\gamma\gamma}[s=2E_{0}hc/\lambda x(1+z)],
\label{tau}
\end{equation}
where $u_{\nu}(z)$ is the co-moving energy density of the photon field, (Stecker, De Jager, \& Salamon ~1992).

In equations (\ref{s}) and  
(\ref{tau}), $E_{0}$ is the observed \gray energy at redshift zero, 
$\lambda$ is the wavelength at
redshift $z$,
$z_{e}$ is the redshift of
the \gray source at emission, $x=(1-\cos\theta)$, \\
$\theta$ being the angle between
the \gray and the soft background photon, 
and the pair production cross section $\sigma_{\gamma\gamma}$ is zero for
center-of-mass energy $\sqrt{s} < 2m_{e}c^{2}$, $m_{e}$ being the electron
mass.  Above this threshold, the pair production cross section is given by

\begin{equation} 
\label{sigma}
\sigma_{\gamma\gamma}(s)=\frac{3}{16}\sigma_{\rm T}(1-\beta^{2})
\left[ 2\beta(\beta^{2}-2)+(3-\beta^{4})\ln\left(\frac{1+\beta}{1-\beta}
\right)\right],
\end{equation} 

\noindent where $\sigma_T$ is the Thompson scattering cross section and $\beta=(1-4m_{e}^{2}c^{4}/s)^{1/2}$  (Jauch \& Rohrlich 1955).

It follows from equation (\ref{s}) that the pair-production cross section has a threshold at $\lambda = 4.75 \ \mu {\rm m} \cdot E_{\gamma}({\rm TeV})$.

The optical depth of the universe to the CBR is given by

\begin{equation}
\tau_{CBR}  = 5.00 \times  10^5 \sqrt{{1.11  PeV}\over {E_{\gamma}}}
\int_0^z  {dz'~(1 +  z') ~{e^{-\left({1.11  PeV}\over  {E_{\gamma}(1 +
z')^2}\right)}}         \over         {\sqrt{\Omega_{\Lambda}        +
\Omega_{m}(1+z')^3}}}
\label{cbr}
\end{equation}
(Stecker 1969; SMS06).

\begin{figure}
\includegraphics[width = 6.5in]{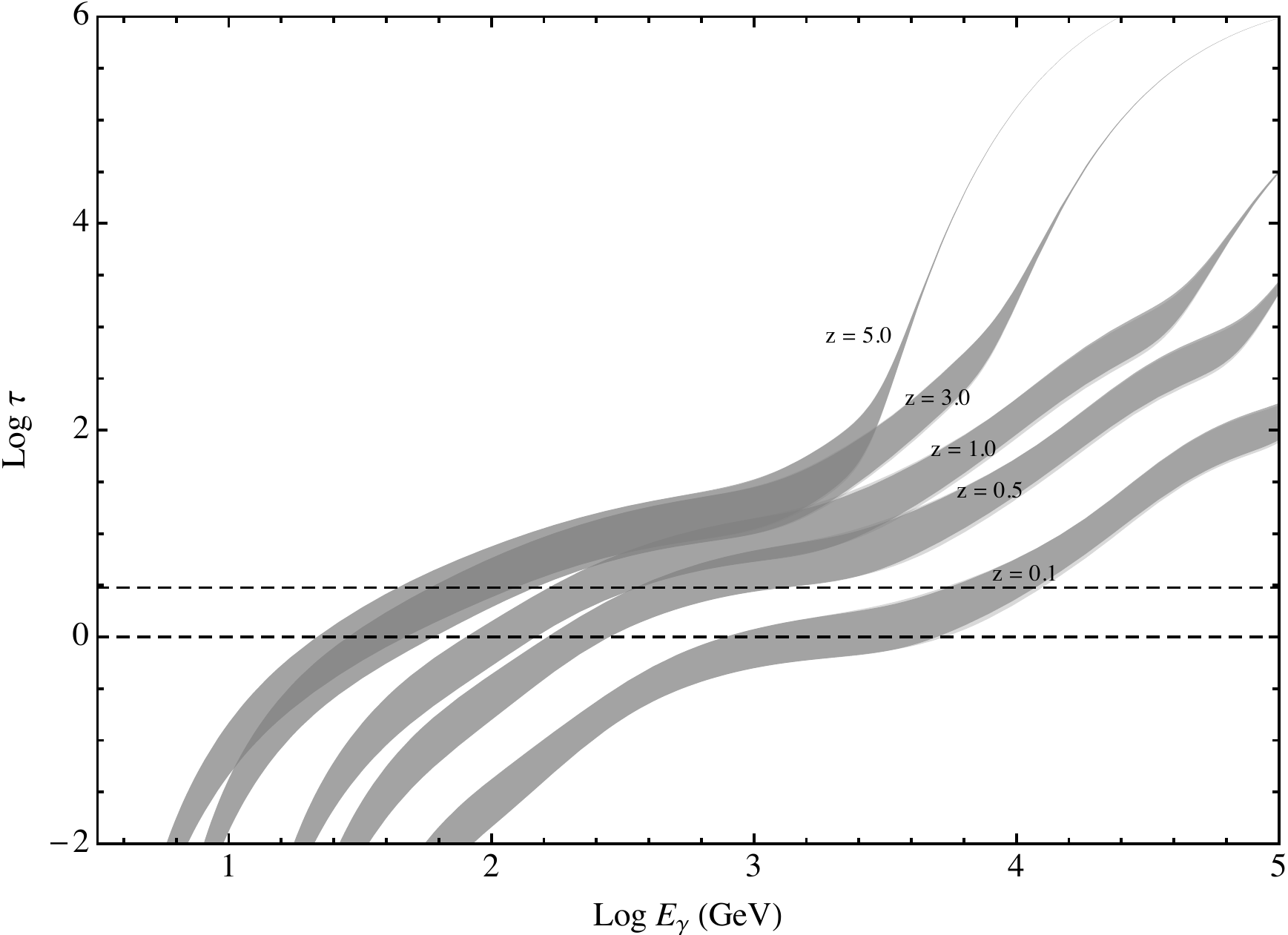}
\caption{The  optical depth  of the  universe from  the  IBL 
and the CBR as well as  the total optical depth as a function of
energy, given for redshifts of 0.1, 0.5, 1, 3, 5. It can be seen that the
contribution  to the  optical depth  from the  IBL dominates  at lower
\gray\ energies and redshifts and that from the CBR photons dominates at the higher
energies and redshifts. The optical depth from CBR photons is an exact function of
energy as given by equation (\ref{cbr}) and therefore the confidence band
is becomes a thin line. The dashed lines indicate the opacities $\tau = 1$ and $\tau = 3$. }
\label{opacity}
\end{figure}

Figure \ref{opacity} shows the 68\% confidence opacity bands for interactions with IBL photons given for sources at $z = 0.1, 0.5, 1, 3 ~$and $5$, calculated using the methods described above, along with the opacity produced by interactions of \grays with photons of the 2.7 K cosmic background radiation. Note that at the higher energies and redshifts where the opacity is dominated by interactions with CRB photons, the uncertainty band becomes a very thin line, since the CBR-dominated opacity is exactly determined by equation (\ref{cbr}).

\section{Discussion and Conclusions}

\subsection{Comparison with our previous backward evolution model}

Ten years ago we made estimates the diffuse infrared background when 
hardly any mid-IR and far-IR luminosity functions
had been observationally determined at high redshifts (SMS06).
That work was therefore based on the assumptions of a
``backward evolution" model. Starting from the well-determined local (z=0) LF
at 60$\mu$m, we assumed that the average locally determined
transformations between different mid-IR and far-IR
wavelengths applied, unchanged, at all redshifts.
The luminosity function used was a double power-law, similar to that
of equation (\ref{lumfunc}) with
the parameters $a = -1.35, b = -3.6$ and
$L_{*} = 8.5 \times 10^{23}$ W/Hz at $z = 0$ as determined at 60 $\mu$m, 
which was the wavelength for which the most complete galaxy LF existed.
We then assumed that the effect of redshift evolution could be
taken account purely by the evolution of $L_{*}$, as described in
SMS06.
  
We have compared the predictions of the SMS06 backwards evolution model
with the recently observed IR LFs as used in this paper. If we make relatively small
improvements in the assumed LF parameters, the backwards evolution
model agrees with the new observations surprisingly well.
The data favor a slightly flatter LF, with a low-luminosity slope
of $a = 1.25$ and a high-luminosity slope of $b = 3.25$, provided that we compensate
by slightly decreasing the normalization of the LF at $L_{*}$ and $z = 0$ to $3.5 \times 10^{-3}$ Mpc$^{-3}$ dex$^{-1}$. We then match the observed LFs at higher redshifts, by assuming that $L_{*}$ increases
as $(1+z)^{3.0}$ for $0 < z < 2.0$, with $L_{*}$ constant at higher redshifts. In this way, we can obtain a good fit between the backward evolution model
and the observational data.

Although this slightly modified backwards evolution reproduces the
observed LFs at all redshifts  well, there are a few 
discrepancies, in either direction.
The most substantial disagreement is with the 8\mic observations of
Huang et al. (2007) at $z = 0.15$.
Below the knee of the LF the observed LF the data
are up to 0.3 dex higher than the LF of our backwards evolution model.
On the other hand, our best-fitting model overpredicts the  15 \mic
LFs by up to 0.25 dex at luminosities above the knee as compared with
the data of Pozzi et al. (2004), Le Floch et al (2005), and Rodighiero et al. (2010).
The likely explanation of both of these discrepancies is that our previously proposed
simple model SEDs are based on average 12\mic luminosities as
derived from the 60 \mic observations. Therefore, they do not include the strong contributions
from the PAH bands (See section 2.2). 
Thus our SEDs will overpredict the dust continuum on either side of
the PAH  bump feature as shown in Figure \ref{PAH}. Correspondingly, if we lower the normalization to better
fit the shorter and longer wavelengths, we under-estimate the
broadband fluxes at 8\mic rest wavelength. 

Of course, our new observationally-based calculation presented in this paper avoids the problems of backward evolution models, since here we use directly observed LFs at 8, 15, and 24 $\mu$m. 
These data include the effect of the PAH emission features and show their significance. 

\subsection{Our present results and comparison with other work}

\begin{figure}
\includegraphics[width = 6.5in]{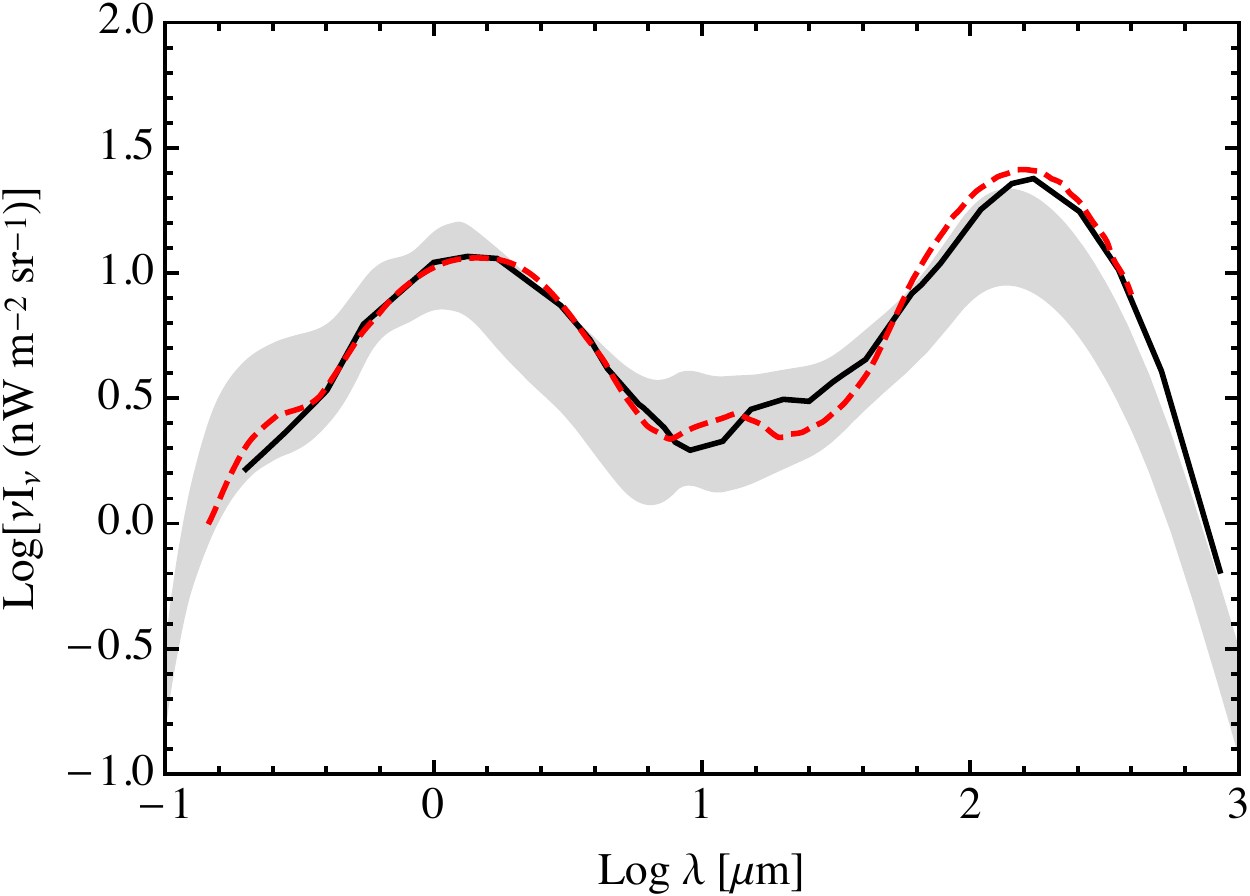}
\caption{A comparison of our confidence band with the models of Franceschini et al. (2008) (solid black line) and Dom\'{i}nguez et al. (2011) (red dashed line).}
\label{ebl-comp}
\end{figure}

\begin{figure}
\includegraphics[width = 6.5in]{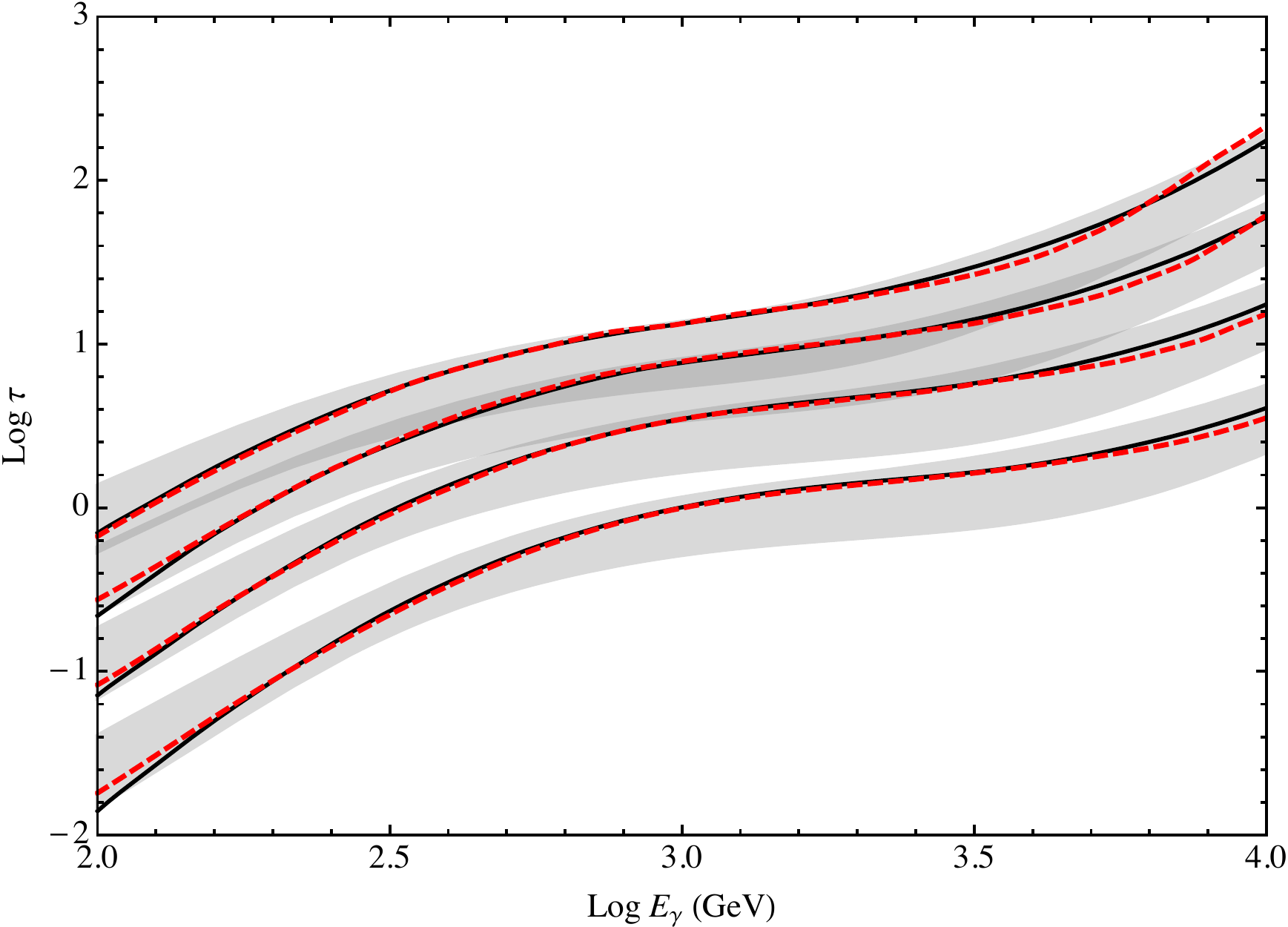}
\caption{Comparison of our opacity results with those obtained by the models of Franceschini et al. (2008) (solid black line) and Dom\'{i}nguez et al.~(2011) (red dashed line).}
\label{opacitycf}
\end{figure}

Figure \ref{ebl-comp} shows our 68\% confidence band computed for the $z = 0$ IBL (EBL) along with EBL SEDs obtained from the models of Franceschini et al. (2008) and Dom\'{i}nguez et al. (2011). 
Figure  \ref{opacitycf} shows our 68\% opacity bands for $z = 0.1, 0.5, 1, 3 ~$and $5$,
in comparison with the opacity curves of Franceschini et al. (2008) and Dom\'{i}nguez et al.~(2011). We note that Franceschini et al. (2008) used data only up to 8 \mic that were available at the time and did not include a PAH component in their model. The model of Dom\'{i}nguez  et al.~(2011) assumes a redshift evolution at redshifts greater than $\sim$ 1 that follows the evolution in the $K$-band given by Cirasuolo et al. (2010). 

\subsection{Conclusion}

In our previous papers (SMS12 and SMS14), we presented observationally based results for the IBL as a function of wavelength and redshift for wavelengths below of 5 \mic. Based on those results, we computed the \gray opacity of the Universe caused by electron-positron pair production up to a \gray energy of $1.6/(1+z)$ TeV. In this paper we have extended our determinations of the IBL within 68\% confidence bands. This determination defines upper and lower limits on the IBL to longer wavelengths that extend to 850 $\mu$m.  Our model-independent results are based on observationally derived luminosity functions from recent galaxy survey data from both local and high redshift surveys. These data include results from {\it Spitzer}, {\it Herschel} and {\it Planck}. We then use these results to calculate the opacity of the Universe to \grays out to PeV energies. In doing so, we also take account of the redshift dependence of interactions of \grays with photons of the 2.7 K cosmic background radiation (CBR) (Stecker 1969), since the opacity from interactions with CBR photons dominates over that from interactions with IBL photons at the higher $\gamma$-ray energies and redshifts. 

\begin{figure}
\includegraphics[width = 6.5in]{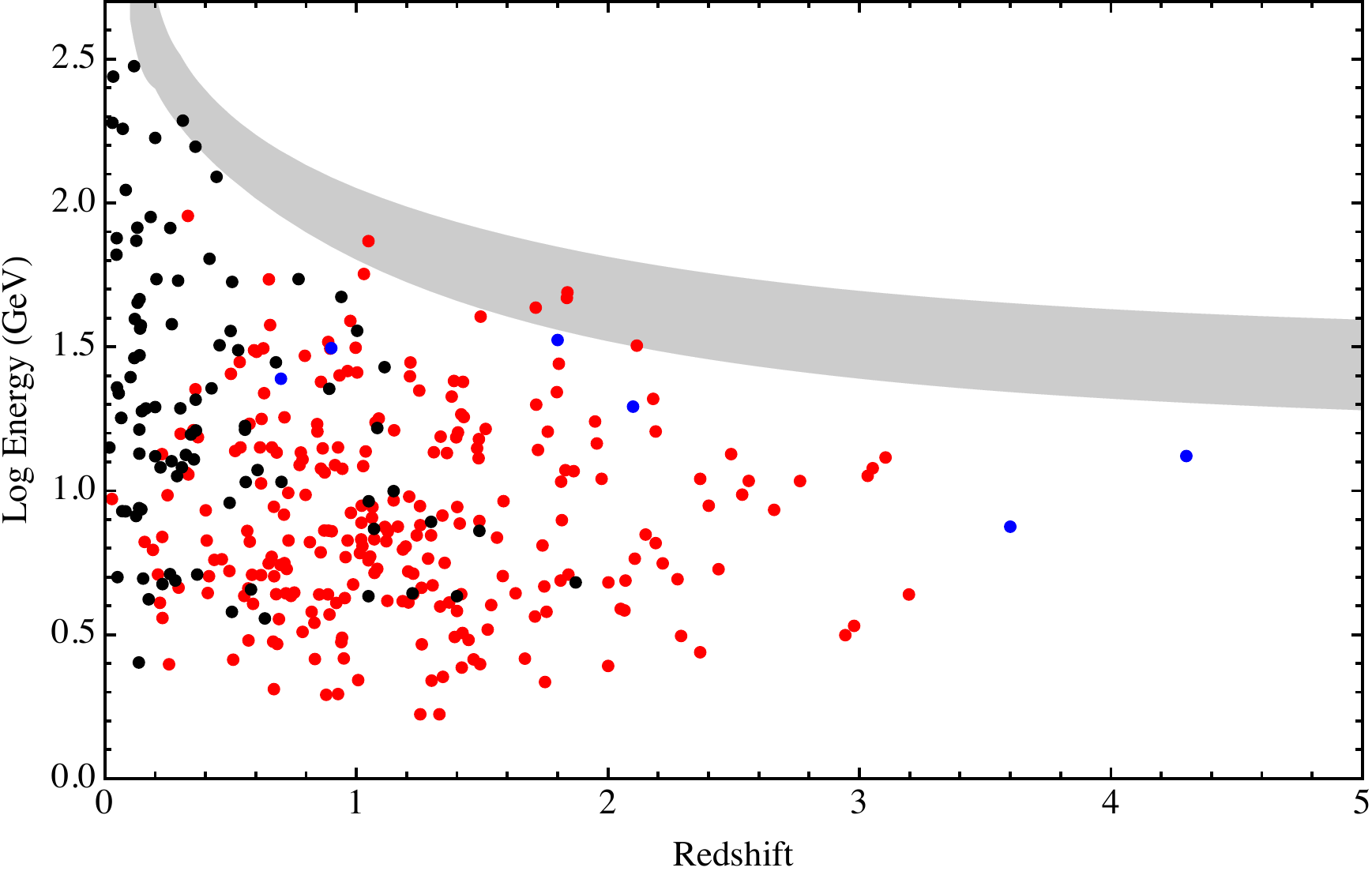}
\caption{A $\tau = 1$ energy-redshift plot (Fazio \& Stecker 1970) showing our uncertainty band results compared with the {\it Fermi} plot of their highest energy photons from FSRQs (red), BL Lacs (black) and and GRBs (blue) {\it vs.} redshift (from Abdo et al. 2010).}
\label{fs}
\end{figure}

In figure \ref{fs} we give an energy-redshift plot showing the highest energy photons from extragalactic sources as a function of redshifts as determined from {\it Fermi} data (Abdo et al. 2010) plotted with our 68\% confidence band for $\tau = 1$.  

\begin{figure}
\includegraphics[width = 6.5in]{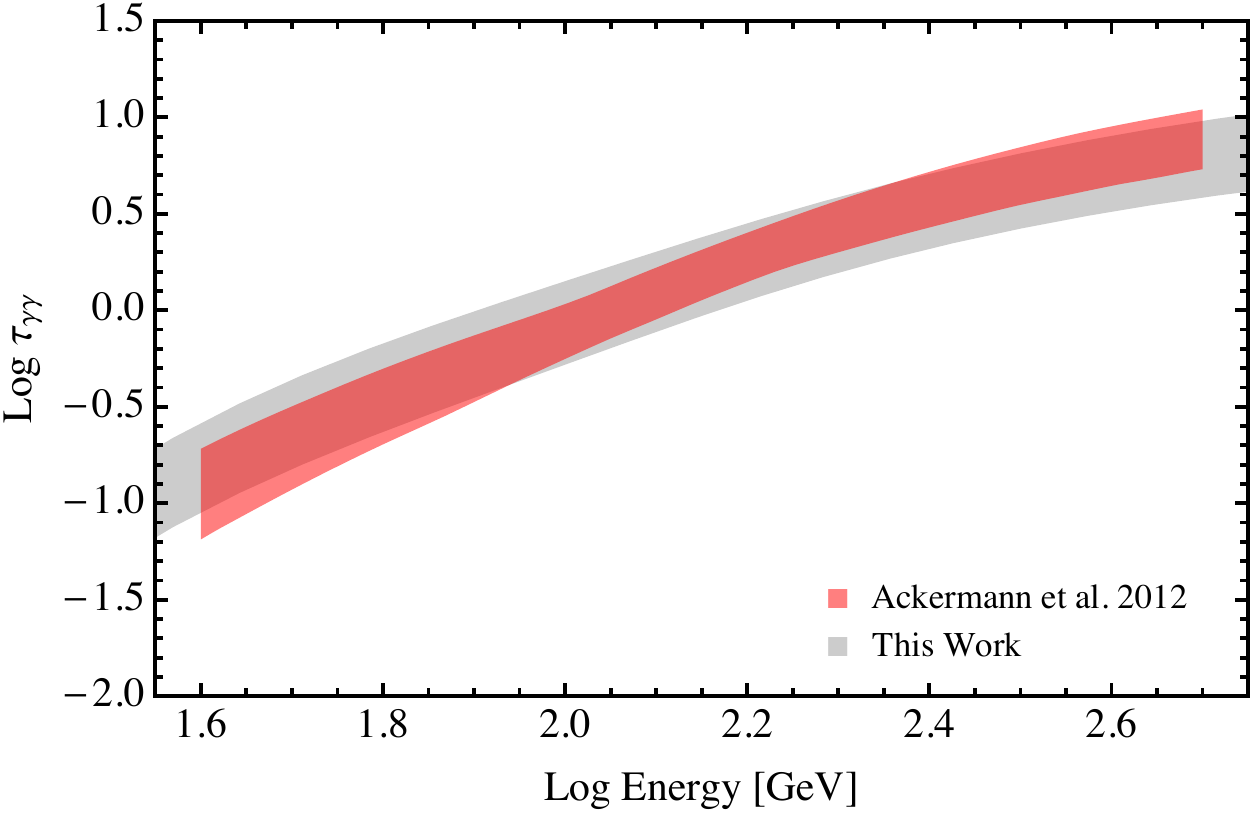}
\caption{Comparison of our results for $z = 1$ with those obtained from an analysis of blazar \gray spectra (Ackermann et al. 2012)}
\label{Ackermann}
\end{figure}

Our direct results on the IBL are consistent with those from complimentary \gray analyses using observations from the {\it Fermi}-LAT \gray space telescope and the H.E.S.S. air \v{C}erenkov telescope. Figure \ref{Ackermann} indicates how well our opacity results for $z = 1$ overlap with those obtained by the {\it Fermi} collaboration (Ackermann et al. 2012). Our results are also compatible with those obtained from higher energy \gray observations using H.E.S.S. (Abramowski et al. 2013). This overlap of results from two completely different methods strengthens confidence that both techniques are indeed complimentary and supports the concept that the spectra of cosmic \gray sources can be used to probe the IBL (Stecker et al. 1992). 

Thus, we find no evidence for modifications of \gray spectra by processes other than absorption by pair production, either by cosmic-ray interactions along the line of sight to the source (Essey \& Kusenko 2014) or line-of-sight photon-axion oscillations during propagation ({\it e.g.}, De Angelis et al. 2007; Mayer \& Horns 2013). In this regard, we note that the {\it Fermi} Collaboration has very recently searched for irregularities in the \gray spectrum of NGC 1275 that would be caused by photon-axion oscillations and reported negative results (Ajello et al. 2016).

We conclude that modification of the high energy \gray spectra of extragalactic sources occurs dominantly by pair production interactions of these \grays with photons of the IBL. They therefore support the concept of using the future {\it \v{C}erenkov Telescope Array} instruments to probe the cosmic background radiation fields at infrared wavelengths. This method can be used in conjunction with future deep galaxy survey observations using the near infrared and mid-infrared instruments aboard the {\it James Webb Space Telescope}.

\newpage

\end{document}